\documentclass[journal=jpccck,manuscript=article]{achemso}

\usepackage[version=3]{mhchem} 

\usepackage{color}
\usepackage{graphicx}
\usepackage{dcolumn}
\usepackage{bm}
\usepackage{multirow}
\usepackage{amsmath}
\usepackage{amssymb}
\usepackage{nameref}

\newcommand{\fref}[1]{Figure~\ref{#1}}

\author{Sagarmoy Mandal}
\affiliation{Department of Chemistry, Indian Institute of Technology Kanpur, Kanpur - 208016, India}
\altaffiliation{Current address: Department of Chemistry, Purdue University, West Lafayette, Indiana 47907, USA}
\author{Tushar Kanti Ghosh}
\email{tkghosh@purdue.edu}
\affiliation{Department of Chemistry, Indian Institute of Technology Kanpur, Kanpur - 208016, India}
\altaffiliation{Current address: Department of Chemistry, Purdue University, West Lafayette, Indiana 47907, USA}

\title{Molecular Insights into the Water Dissociation and Proton Dynamics at the $\beta$--TaON (100)/Water Interface}
\abbreviations{IR,NMR,UV}

\begin{document}







\begin{abstract}
  Understanding the dynamic nature of the semiconductor-water interface is crucial for developing efficient photoelectrochemical water splitting catalysts, as it governs reactivity through charge and mass transport.
In this study, we employ {\em ab initio} molecular dynamics simulations to investigate the structural and dynamical properties of water at the $\beta$-TaON (100) surface.
We observed that a well-defined interface is established through the spontaneous dissociation of water and the reorganization of surface chemical bonds.
This leads to the formation of a partially hydroxylated surface, accompanied by a strong network of hydrogen bonds at the TaON-water interface.
%
Consequently, various  proton transport routes, including the proton transfer through ``low-barrier hydrogen bond'' path, become active across the interface, dramatically increasing the overall rate of the proton hopping at the interface.
Based on our findings, we propose that the observed high photocatalytic activity of TaON-based semiconductors could be attributed to the spontaneous water dissociation and the resulting high proton transfer rate at the interface.
\end{abstract}

\section{Introduction}
TaON is a promising photoactive material for water splitting, capable of functioning under visible light due to its appropriate band gap and band positions.~\cite{Hitoki:ChemCommun:2002,chen_jacs:2023,Chun:JPCB:2003,Maeda:JPCC:2007,Yashima:ChemMat:2007,Zhang:CSR:2014,reshak_pccp:2014}
Due to this, there has been significant attention given recently to understand the structure, morphology, and efficiency of TaON-based materials under experimental conditions, with the aim of making future improvements.~\cite{jouypazadeh_ass:2021, jouypazadeh2021dft,Xiao_ACR:2023,Xiao_AngewChem:2022}
In addition to the basic requirements of band gap and band positions, the efficiency of overall water splitting on TaON depends on the mechanisms, thermodynamics, and kinetics of photochemical reactions at the TaON-water interface.~\cite{Linsebigler:ChemRev:1995,Lewis:ChemRev:2010,Kudo:CSR:2009,Maeda:JPCL:2010,Hisatomi:ChemSocRev:2014,Maeda:AcsCat:2013,Maeda:JPCC:2007,Osterloh:CSR:2013,catal_Nadeem:2021,Nishioka_NatRev2023}
This includes considerations of charge and mass transfer across the interface~\cite{Xiong_CatalSciTech:2014,ChemSci_Minegishi:2013}, especially when TaON is used as a catalyst for water splitting in suspended water.~\cite{Ito_PCCP:2005}
The interface formed by water and TaON plays a crucial role in controlling the mechanisms of water splitting reactions that occur at the water-TaON hetero-junction.
This is because the photo-catalytic oxidation and reduction reactions on TaON are believed to proceed through the migration of photo-generated hole ($h^{\rm +}$) and electron ($e^{\rm -}$) produced in the bulk to the active sites on its surfaces, where water molecules and co-catalysts are adsorbed, followed by surface chemical reactions.~\cite{Linsebigler:ChemRev:1995,Hisatomi:ChemSocRev:2014}
The activity depends on the local chemical environments formed at the interface through the reorganisation of chemical bonds.
These rearrangements produce new active species on the catalyst surface.
The efficiency of the overall process crucially depends on the thermodynamic stability and kinetic lability of these active chemical species, such as protons, hydroxyl ion, and hydronium ion, at the interface.
Several studies have revealed that water dissociation and proton transfer at the interface can influence the catalytic reactions on the
surface.~\cite{Andersson:JACS:2008,Moilanen:Langmuir:2008,Sheng:AngewChem:2013,acsomega_2022} 
These processes play an even larger role in hydrogen evolution reactions during photochemical water splitting using semiconductors.
Wood {\em {et. al}}~\cite{Wood:JACS:2013} showed that increased proton transfer at the InP--water interface enhances the hydrogen evolution activity when using this semiconductor with Pt co-catalyst. 
They attributed this enhancement to the formation of ice-like structures at the interface due to strong hydrogen bonding between water molecules, amplifying the proton transfer process.
A higher proton transfer rate reduces the barrier for H$_2$ evolution reactions, favoring overall water splitting.
The importance of proton transfer in photochemical water splitting at the (10$\bar 1$0) GaN--water interface
 has also been reported.~\cite{Wang:JPCC:2012}
Other examples include rapid, long-range proton and hydroxide transfer at Ceria-water interface,\cite{Tateyama_jacs:2016} solvation-induced proton diffusion at (10$\bar 1$0) ZnO--water interface,\cite{Tocci:JPCLett:2014} and the generation of new active sites due to solvent polarization effect at Au/TiO$_2$--water interface.\cite{Farnesi:JPCL:2013}
Studies on the activity of water--ZrO$_2$ interface~\cite{yang_jpcc:2021}, and investigations into one dimenstional versus two dimensional proton transport in ZnO\cite{matti_CS:2019} have also been conducted.
Additionally, it has been reported that the diffusion of protons on the FeO thin film is entirely governed by a water-mediated hopping mechanism.\cite{merte_sci:2012}

Despite the significant attention given to understand the structure, composition and morphology to improve the catalytic activity of TaON-based materials~\cite{huang_APA:2023,Ullah_ACBE:2018,Taviot_jpcc:2015,Narayanachari_ACSaem:2020}, little focus has been placed on comprehending the kinetics of the processes occurring during the photochemical water splitting activity of these materials. 
This is primarily because understanding the formation of the interface itself is a highly intriguing process for experimental study,\cite{barry_chemrev:2021,merte_sci:2012} and unraveling the mechanisms of interface assisted chemical processes introduces additional complexities.~\cite{Ito_PCCP:2005}
Quantum chemical calculations have been used in the recent past to understand the properties of TaON based materials.\cite{Ullah_ACBE:2018,Nair_TaON:SurfSci:2014,lahmer_ccm:2023,Esch_RRL:2014,lahmer_ccm:2022,reshak_pccp:2014}
However, most of these calculations have focused on understanding the electronic and band structure of TaON based materials using static approach, which does not provide detailed insights into the dynamic behavior of the interface.
On the other hand, {\it ab initio} molecular dynamics (AIMD), a powerful technique for investigating chemically complex systems with intricate structural and dynamical information~\cite{pham_natmat:2017,english_pccp:2014,barry_chemrev:2021,Blumberger_jpcl:2018,Blumberger_pccp:2022,vonRudorff_2016,Tateyama_jacs:2016,Tateyama_jpcc:2010,Tateyama_jpcc:2015,Tateyama_jpcc:2020,jpcl_behler,jctc_sprik}, can be employed to study such dynamic processes at the TaON-water interface. 
The potentials of AIMD techniques in studying and designing new water splitting catalysts have been reviewed in recent literature.~\cite{Goga_catal:2021}

In this paper, we have conducted a study of the structure and dynamics of water at the $\beta$--TaON (100) surface by employing density functional theory (DFT) based AIMD simulation.  
In previous studies, it was demonstrated that the $\beta$--TaON (100) surface is the most stable among all surfaces.~\cite{Nair_TaON:SurfSci:2014,Esch_RRL:2014,Narayanachari_ACSaem:2020}
This surface possesses Lewis acid-base sites due to the presence of under-coordinated surface Ta and O atoms,  making them reactive toward water dissociation. 
In this work, our primary focus is on understanding the structure and the dynamics of water, particularly water dissociation and proton transfer on the (100) $\beta$--TaON surface. 
By employing AIMD simulation, we investigated the rate, mechanisms and free energy barrier of water dissociation, as well as the structure and hydrogen bonding network formation at the interface. 
To the best of our knowledge, our study represents the first in-depth molecular-level analysis of the structural and dynamic aspects of the TaON-water interface.

\section{Methods and Model}
\begin{figure}[t]
\begin{center}%
\includegraphics[scale=0.55]{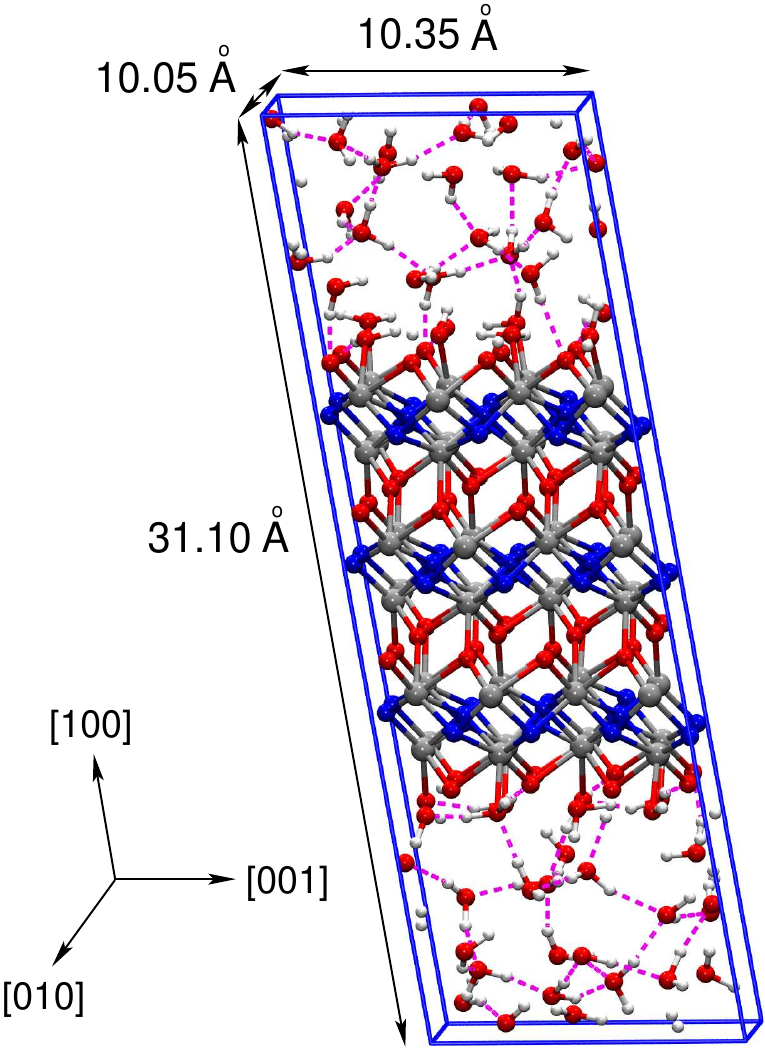}
\end{center}%
\caption[]{%
A snapshot of (100) $\beta$--TaON simulation cell used in our calculation. 
Atomic color code: silver (Ta), blue (N), red (O)  and white (H). The dotted magenta lines represent the 
hydrogen bonds (HBs). 
HBs are defined based on the following geometric conditions: distance
$d$[O--O] {$\leqslant$} 3.5 {\AA} and angle $\theta$[O--H$\cdots$O] {$\geqslant$} 140$^{\rm o}$. 
}
\label{Sim_cell}%
\end{figure}

In our study, we used the conventional periodic slab model for simulating the solid surface.
A slab of 3$\times$2$\times$2 Ta$_{48}$O$_{48}$N$_{48}$ supercell, based on experimental lattice parameters,~\cite{Li:InorgChem:2010} was used for modelling the (100) $\beta$--TaON surface.
%
The slab was chosen in such a way that the top and the bottom layers are equivalent and symmetric with respect to the center of inversion,
%
%
avoiding any unphysical dipolar interaction between the slabs.
Our previous benchmark calculations showed the importance of such supercell in converging the surface properties 
with respect to surface unit cell size, k-points, and slab thickness.~\cite{Nair_TaON:SurfSci:2014}
%
For our present simulation, we chose the surface termination that was identified as the most stable in our previous study.~\cite{Nair_TaON:SurfSci:2014}
%
%
In this surface termination, the clean (100) surface contains six-fold coordinated Ta (Ta$_{\rm 6c}$) atoms along with two-fold coordinated O (O$_{\rm 2c}$) atoms.
$\beta$--TaON/water interface was created by filling the space between 
two periodically repeated slabs with water, along the direction normal to the surface. 
A liquid water slab of 16~{\AA} thickness, containing 59 water molecules, was included in the space between two TaON surfaces.
The number of water molecules were taken such that the water density
is that of the bulk water (0.99 g~cm$^{\rm -3}$) at 300~K.
%
%
Overall, a supercell of 31.10 $\times$ 10.05 $\times$ 10.35~{\AA}$^{\rm 3}$ with $\beta$ = 99.64$^{\circ}$ 
was used in our simulation.
A snapshot of the supercell is shown in~\fref{Sim_cell}.

All the computations were performed using periodic
DFT employing
the {\tt CPMD} program~\cite{cpmd}.
We employed the PBE~\cite{pbe} exchange correlation functional together with
ultrasoft--pseudopotentials~\cite{vanderbilt} and a plane--wave cutoff of 30~Ry.
Canonical ensemble Car--Parrinello molecular dynamics (CPMD)~\cite{marx-hutter-book} simulations were carried out using the Nos{\'e}--Hoover chain thermostat~\cite{nhc}
at 300~K with a time step of 5.0~a.u.
Fictitious masses for orbitals were taken as 700 a.u. and hydrogen masses were replaced by deuterium masses.
%
%
A trajectory with a total length of 67~ps was generated, and for the analysis of equilibrium properties, the final 20 ps of the trajectory were considered.

\section{Results and Discussion}
\subsection{Formation of (100) TaON--Water Interface }
\label{form_int}
\begin{figure}[!ht]
\begin{center}
\includegraphics[scale=0.85]{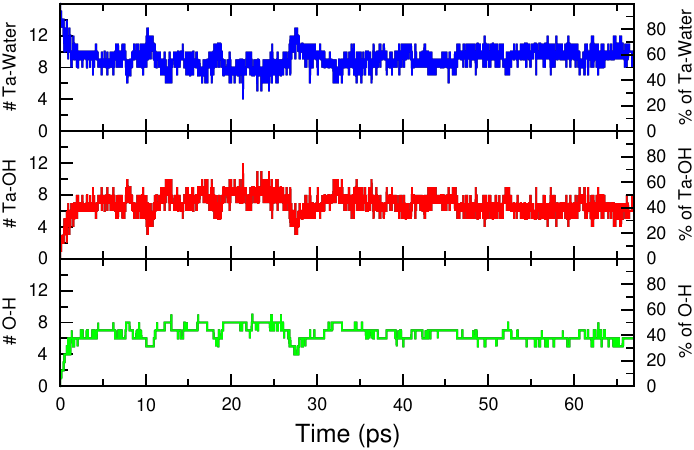}
\end{center}
\caption[]{%
Fluctuations in the total number (as well as the percentage) of surface chemical bonds as a function of time during the canonical--ensemble MD simulation at 300~K.
A cutoff distance of 2.6 {\AA} was used to determine chemical bonds between surface Ta and water or OH$^{\rm -}$.
We considered O and H atoms to be bonded when the distance between them is less than 1.2 {\AA}.
}
\label{surf_100}
\end{figure}

%
We started our MD simulation with a configuration where clean non-hydroxylated (100) TaON 
surfaces were made in contact with the liquid water slab.
Upon exposure to water molecules, the top and bottom surfaces readily reacted with water molecules to form a well characterizable interface structure. 
%
%
All 16 Ta$_{\rm 6c}$ atoms (including top and bottom surfaces), acting as Lewis acid centers on the surface, bonded with the incoming water molecules to initiate the reorganisation of the surface chemical bonds.
Some of the adsorbed water molecules dissociated on the surface and donated protons to O$_{2c}$ atoms, leading to the hydroxylation of the surface and creation of dissociated ({\bf D})-type waters.
%
Meanwhile, rest of the absorbed water molecules remained in the molecular ({\bf M}) form.
These processes are spontaneous and primarily responsible for the generation of active intermediates, such as surface O--H, Ta-OH and Ta-water species.
Consequently, a surface configuration is formed which is similar to the structure observed in previous static DFT calculation with one mono-layer (ML) of water molecules on the surface.~\cite{Nair_TaON:SurfSci:2014}
However, the intrinsic nature of the surface is dynamic and various active species remain on the surface, continuously transitioning from one state to another. 
This dynamic behavior is clearly observed when concentration of these active species were monitored during our MD simulation (see \fref{surf_100}). 
Initially, about 40\% of the surface O atoms were hydroxylated within $\sim$2~ps,
and the fluctuations in the number of surface O--H bonds continued beyond this point. 
Similarly, fluctuations in the number of Ta-water and Ta-OH bonds were observed during the course of our MD simulation, 
%
reflecting the occurrence of various proton transfer events between the active species on the surface.
These observations further confirm the highly reactive nature of the interface.
After the equilibration phase, $\sim$40\% of the surface adsorbed waters were dissociated, resulting in the protonation of $\sim$40\% of the surface O atoms.

The observed rapid surface protonation can be explained by the electronic structure of the clean (100) $\beta$--TaON surface.~\cite{Nair_TaON:SurfSci:2014}
In the projected density of states, the contribution of the O (O$_{\rm 2c}$) atoms near the edge of the valence band makes the surface basic in nature and accelerates the proton transfer from the adsorbed water to these surface oxygen atoms. 
However, the basicity of these O$_{\rm 2c}$ atoms is not strong enough to hold the proton for a very long time.
Therefore, O$_{\rm 2c}$ atoms donate the proton back to the water molecules, making the process reversible.

\subsection{Rate and Pathways of Proton Transfer}
\label{pt_rate}

\begin{figure*}[ht]
\begin{center}
\includegraphics[scale=0.5]{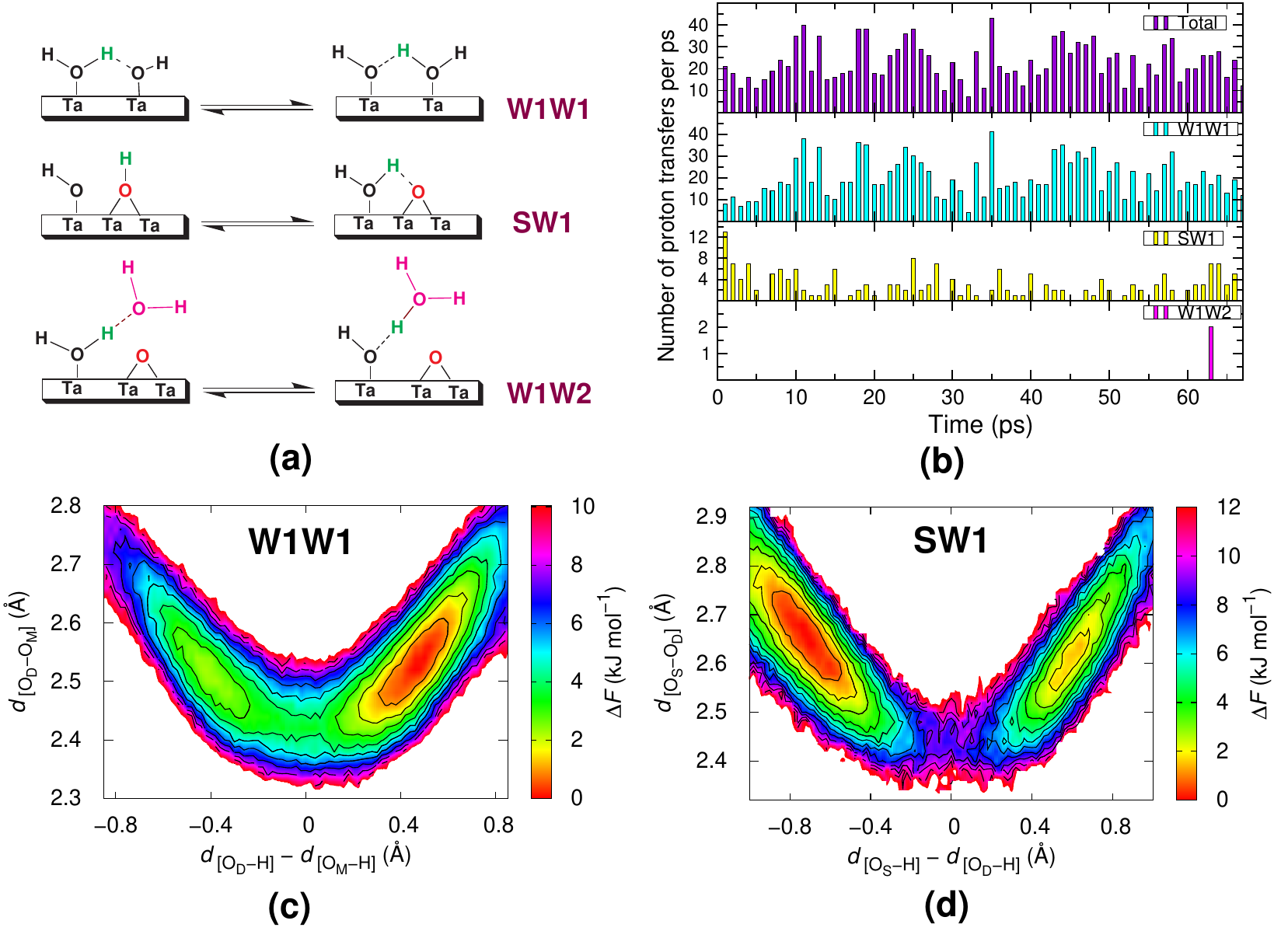}
\end{center}
\caption[]{%
(a) Various mechanistic routes of proton transfers are illustrated. Surface O$_{\rm{2c}}$ atoms are indicated in red and water molecules in 
the second adsorption layer are represented in magenta. The protons participating in the hopping process are indicated with green color.
(b) The rate of proton transfer through various mechanistic routes during the MD simulation.
``Total'' shows the overall number of proton transfers per ps through all the mechanistic routes.
(c) Free energy surface for the proton transfer through {\bf W1W1} mechanism. 
(d) Free energy surface for the proton transfer through {\bf SW1} mechanism. The contour lines are separated by 1 kJ/mol energy.
The surface O atom is denoted by O$_{\rm S}$ and O atoms of {\bf D}-type and {\bf M}-type waters are denoted by O$_{\rm D}$ and O$_{\rm M}$, respectively.
}
\label{proton_transfer}
\end{figure*}

The dynamical behavior of the reactive species at the interface give rise to various active proton hopping routes (see \fref{proton_transfer}(a)),
%
namely (1) {\bf W1W1}: proton transfer between {\bf M}-type water on a Ta atom and {\bf D}-type water on a neighboring Ta atom, 
(2) {\bf SW1}: proton transfer between {\bf D}-type water on a Ta and a surface O--H group,
and (3) {\bf W1W2}: proton transfer between {\bf M}-type water on a Ta atom and a water molecule in the second adsorption layer. 
The rate of proton transfer across the surface through these mechanisms was estimated by calculating the number of proton transfer events per picosecond of the trajectory (see ~\fref{proton_transfer}(b)).
%
Our analysis revealed an average rate of 22 proton transfers per picosecond during the last 10 ps of the simulation, indicating a significantly high rate for a surface with only 16 Ta atoms.
The dominant pathways for proton hopping were found to be {\bf W1W1} and {\bf SW1}, which remained active throughout the entire simulation. 
In contrast, proton transfer through the {\bf W1W2} mechanism was rare and occurred only once in the last few ps of the MD run. 
%
Notably, the {\bf W1W2} proton transfer pathway led to the formation of H$_{\rm 3}$O$^{\rm +}$ ion in the second adsorption ({\bf W2}) layer via the abstraction of a proton from the first adsorption ({\bf W1}) layer.
However, this H$_{\rm 3}$O$^{\rm +}$ intermediate was short-lived and readily deprotonated to the hydroxyl ion of the {\bf W1} layer.
Interestingly, the {\bf W2} layer exhibited minimal participation in proton hopping events, primarily due to its spatial separation from the {\bf W1} layer, as explained in later sections.

\subsection{Free Energy Barriers of Proton Transfer}
\label{pt_fes}

To understand the underlying factors contributing to the high rate of proton transfer at the interface, we aimed to estimate the free energy barriers associated with the two most frequently observed mechanisms,
i.e., {\bf W1W1} and {\bf SW1}. 
The free energy surfaces were generated using the standard equation: $$\Delta F=-k_{\rm B}T\ln P(d_{[\rm O_{1}-O_{2}]}, d_{[\rm O_{1}-H]}-d_{[\rm O_{2}-H]}).$$
 Here, $P(d_{[\rm O_{1}-O_{2}]}, d_{[\rm O_{1}-H]}-d_{[\rm O_{2}-H]})$ is the two-dimensional probability distribution function of the standard proton transfer coordinates, $d_{[\rm O_{1}-O_{2}]}$ and $d_{[\rm O_{1}-H]}-d_{[\rm O_{2}-H]}$.
$d_{[\rm O_{1}-O_{2}]}$ is the distance between O$_{\rm 1}$ and O$_{\rm 2}$ atoms
involved in proton hopping.
The distances of the H atom (participating in the hopping) from the O$_{\rm 1}$ and O$_{\rm 2}$ atoms are $d_{[\rm O_{1}-H]}$ and $d_{[\rm O_{2}-H]}$, respectively.
The free energy surface for {\bf W1W1} mechanism is shown in \fref{proton_transfer}(c). 
In this case, the proton hops between a molecularly adsorbed {\bf M}-type water and a neighboring dissociatively adsorbed {\bf D}-type water molecule.
%
%
The forward and reverse free energy barriers of proton transfer are $\sim$2 kJ/mol and $\sim$4 kJ/mol, respectively.
This indicates that the proton is attached to the {\bf M}-type water most of the time compared to the {\bf D}-type water, and thus the minimum on the right side is deeper.
%
%
The next prominent proton hopping mechanism is {\bf SW1}, where 
proton hops between a dissociatively adsorbed {\bf D}-type water and a surface O--H group.
The free energy surface for {\bf SW1} mechanism is shown in \fref{proton_transfer}(d).
The minimum on the left side of the free energy surface corresponds to the state where the proton is attached to the surface O atom (O$_{\rm S}$),
whereas the shallow right side minimum corresponds to the state where proton is attached to the O atom of dissociatively adsorbed water (O$_{\rm D}$). 
The free energy barrier for the forward path is $\sim$9 kJ/mol,
while the reverse path has a free energy barrier of $\sim$8 kJ/mol.
This indicates that water dissociation to the surface O atom is kinetically favorable at room temperature. Our previous calculation with ML of water also shows that such a partially hydroxylated surface is thermodynamically stable.\cite{Nair_TaON:SurfSci:2014}
%
%
The barrier for the proton transfer is found to be highly sensitive to the distance between the two oxygen atoms that are directly involved in the proton transfer process.
The average $d_{[\rm O-O]}$ is 2.6~{\AA} for the {\bf W1W1} mechanism, while for {\bf SW1} mechanism, average $d_{[\rm O-O]}$ is 2.7~{\AA}.
Such d[O-O] distances facilitate the formation of low-barrier hydrogen bonds (HBs)~\cite{Birgit:PNAS:1998} between the O atoms, contributing to the high rate of proton transfer at the interface.
Similar phenomena have been observed at various semiconductor-water interfaces, such as
the (10$\bar 1$0) GaN--water interface with a reported barrier\cite{Wang:JPCC:2012} of 40 meV and an average $d_{[\rm O-O]}$ of 2.5 {\AA}, as well as
the (10$\bar 1$0) ZnO--water interface with a barrier\cite{Tocci:JPCLett:2014} of 70 meV and $d_{[\rm O-O]}$ of 2.4 {\AA}.

\subsection{Density Profile of Atoms Normal to the Surface}
\label{atom_density}

\begin{figure*}[!ht]
\begin{center}
\includegraphics[scale=0.55]{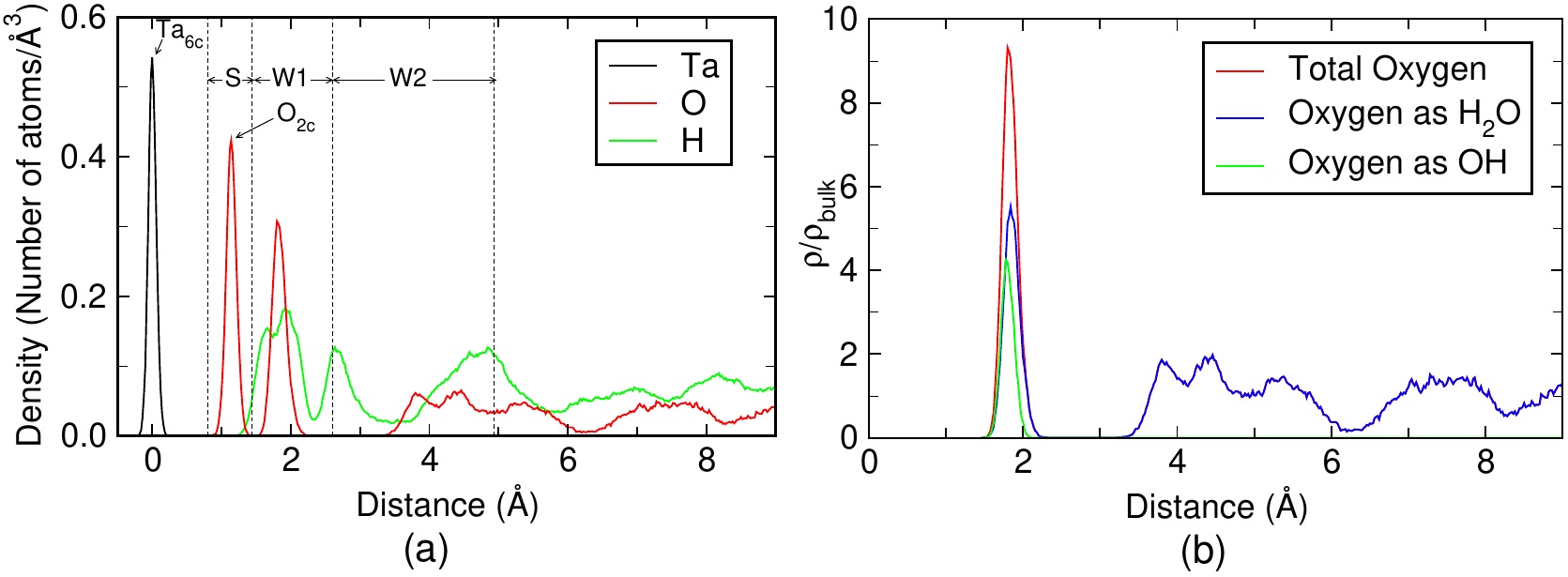} 
\end{center}
\caption[]{%
(a) Planer averaged number densities of atoms at the interface as a function of distance along the surface normal.
For calculating the distances, the plane formed by the eight equivalent Ta$_{\rm 6c}$ surface atoms was used as the reference plane.
The different regions formed at the interface are labeled as {\bf S} (surface layer), {\bf W1} (first adsorbed water layer), and {\bf W2} (second adsorbed water layer), with
boundaries indicated by vertical dotted lines.
(b) Planer averaged relative density of oxygen atoms at the interface as a function of distance along the surface normal.
%
%
The decomposition of total O density, based on the number of protons attached to the oxygen atom, is also shown.
%
}
\label{density_profile}
\end{figure*}

In the previous sections, our focus was primarily on understanding the dynamic processes occurring at the interface and their associated mechanisms.
Now, we shift our attention to the equilibrium properties of the interface, specifically exploring the impact of these dynamic processes on the interface structure, average spatial orientation of water molecules, and the formation of the HB network.

%
%

To gain a deeper understanding of the interface structure, 
we analyzed the 
planar averaged density profile of different atoms as a function of their distance from the surface (see~\fref{density_profile}(a)). 
%
%
The density profile of oxygen and hydrogen atoms shows several distinct peaks at varying distances from the surface, indicating the formation of a well defined layered structure at the interface.
%
%
Notably, the sharp peak of O atom at 1.14 {\AA} corresponds to a layer formed by the surface O (O$_{\rm 2c}$) atoms.
Similarly, another sharp peak of O 
at 1.80 {\AA} is attributed to all the O atoms of water molecules that are immediately adsorbed (molecularly or dissociatively) on the surface and directly bonded to the surface Ta (Ta$_{\rm 6c}$) atoms. 
%
%
The double peak of O around 4 {\AA} indicates the presence of a second adsorbed layer of water molecules.
However, these water molecules are more flexible and form a broader peak compared to the previous layers, indicating a higher degree of mobility. 
%
%
%
%
In order to differentiate these layers, the density profile is divided into three distinct regions: {\bf S} (surface layer), {\bf W1} (first adsorbed water layer), and {\bf W2} (second adsorbed water layer), as shown in~\fref{density_profile}(a).
%
%
%

To identify the protonated or deprotonated state of the water molecules at interface, the planer averaged relative density profile of O atoms and its decomposition based on the number of attached protons are shown in \fref{density_profile}(b).
%
The plots reveal that the {\bf W1} layer consists of both dissociated and molecular water, giving rise to sharp peaks for both OH and H$_2$O groups within the {\bf W1} region (at around 1.8 {\AA} from the surface).
The pronounced peak in the total O density within the {\bf W1} layer indicates a significantly increased water density at
the TaON/water interface.
%
In contrast, no dissociated water molecules were found beyond 2.5 {\AA} from the surface, suggesting that all water molecules in the {\bf W2} layer and above remain in the molecular form.
%
Interestingly, water molecules in the {\bf W2} layer give rise to two peaks at $\sim$3.80 {\AA} and $\sim$4.46 {\AA}.
This is due to the fact that the water molecules in the second layer are positioned at two different distances from the surface in order to form stable HB network with {\bf S} and {\bf W1} layer. 
%
%
During the formation of hydrogen bonding network, the first and second peaks of the {\bf W2} layer maintain a distance of $\sim$2.66 {\AA} from the peaks of {\bf S} and {\bf W1} layers, respectively.
%
%
Consequently, the water density declines to zero in between the {\bf W1} and {\bf W2} regions.

\begin{figure}[t]
\begin{center}
\includegraphics[scale=0.370]{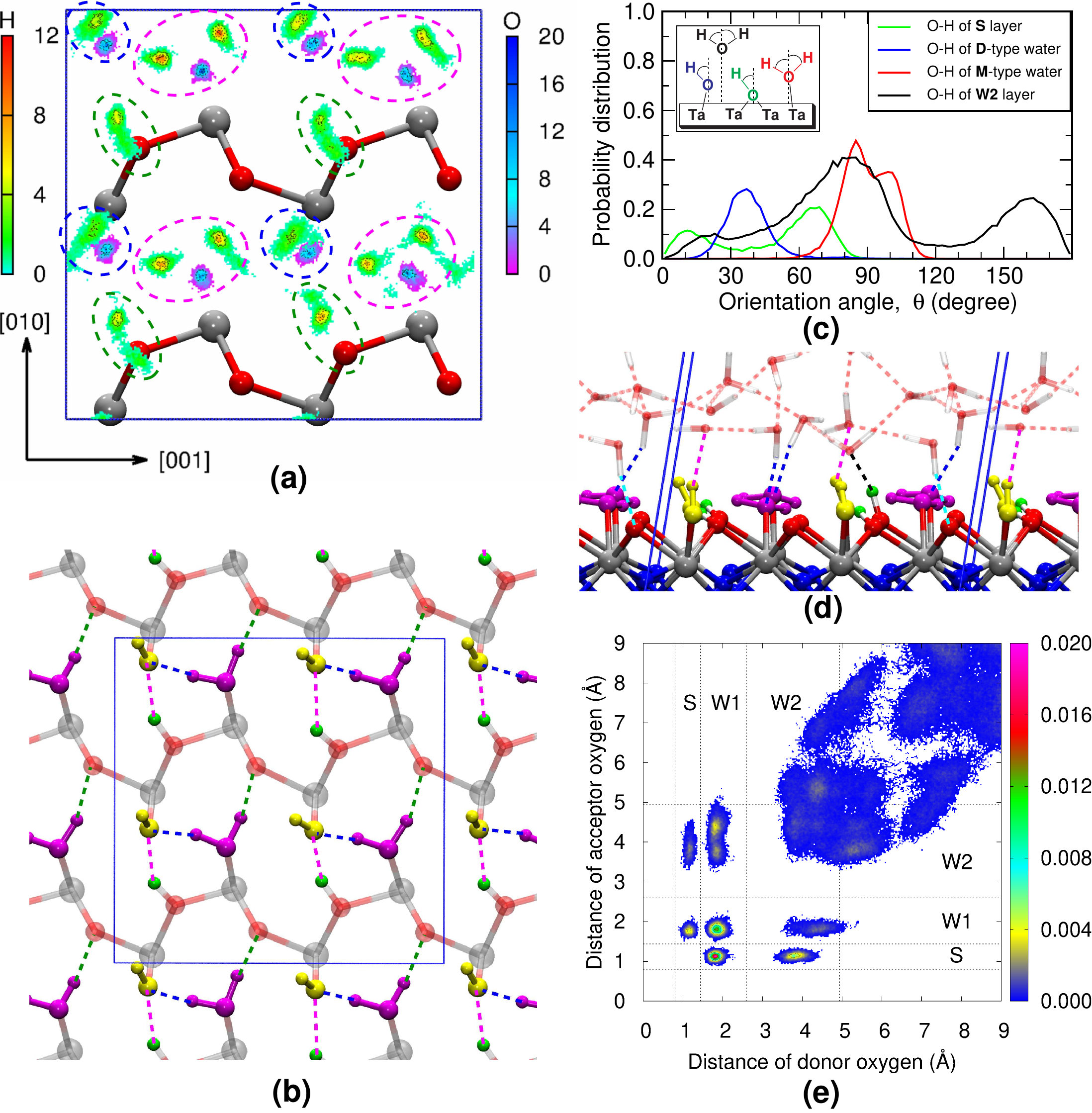}
\end{center}
\caption[]{%
(a) Spatial probability distribution of O and H atoms of natively adsorbed water molecules projected on (100) surface. 
The surface Ta and O atoms are shown in ball and stick model. 
%
{\bf D}-type water, {\bf M}-type water,
and H atoms adsorbed to the surface are encircled with blue, magenta, and green ovals, respectively
(b) Top view of a representative structure of natively adsorbed water molecules on the surface.
%
{\bf D}-type and {\bf M}-type waters are represented in yellow and magenta, while the H atoms adsorbed to the surface are shown in green.
HBs of the {\bf M}--{\bf D}, {\bf M}--{\bf S} and {\bf S}--{\bf D} types are indicated by blue, green, and magenta dashed lines, respectively.
(c) Distribution of the orientation angle of O-H bonds with respect to the surface normal direction.
The definition of orientation angle, denoted as $\theta$, is shown in the inset.
(d) Side view of the TaON--water interface.
Water molecules of {\bf W2} and upper layers are shown in transparent colors. 
HBs within {\bf W2} and bulk layers are represented by red dashed lines.
%
%
HBs of {\bf W2}--{\bf W1}, {\bf W1}--{\bf W2}, {\bf S}--{\bf W2}, and {\bf W2}--{\bf S} types are indicated by blue, magenta, black, and cyan dashed lines, respectively.
(e) Planar averaged probability density of HBs 
as a function of distances of donor O atoms and acceptor O atoms from the surface. 
We categorized the position of donor and acceptor atoms into {\bf S}, {\bf W1}, and {\bf W2} regions based on our previous analysis. 
}
\label{combined}
\end{figure}

\clearpage

\subsection{Spatial Probability Distribution of Surface Adsorbed Waters}
\label{spl_density}
%
To visually depict the preferred positions and orientations of surface adsorbed water molecules,
the spatial probability distribution of the H and O atoms for the covalently bonded water molecules at the surface is shown in~\fref{combined}(a).
%
In the contour plot, the localized spatial distribution of O atoms, close to the surface Ta atoms, corresponds to the chemically bonded waters of {\bf W1} layer. 
The spatial distribution of H atoms appears to be more diffused, with specific patterns indicating the presence of molecular (within magenta oval) and dissociative (within blue oval) water molecules, as well as surface adsorbed protons (within green oval).
A more simplified view of different modes of water adsorption is presented in \fref{combined}(b), illustrating {\bf D}-type waters (yellow), {\bf M}-type waters (magenta) and protons adsorbed to the surface O atoms (green).
The observed arrangement of alternating {\bf D}-type and {\bf M}-type water molecules on the surface along the [001] direction is consistent with our earlier calculation
with ML water coverage.~\cite{Nair_TaON:SurfSci:2014}
Additionally, the appearance of diffused spatial probability distribution of H atoms in between the O atoms of {\bf D}-type and {\bf M}-type waters further supports the occurrence of frequent proton transfer events, as discussed before.
A similar arrangement of dissociated and molecular waters at the surface has also been observed for ZnO (10$\bar 1$0) surface with ML water coverage.~\cite{Tocci:JPCLett:2014}

\subsection{Hydrogen Bonding Network and Orientation of Water  Molecules}
\label{HB_net}
More valuable insights about the solid--water interactions can be obtained by analyzing the hydrogen bonding network present at the interfaces. 
Here, we focus on understanding such HB patterns across different layers based on the planar averaged density of HBs, computed as a function of distances of donor and acceptor O atoms from the surface, as shown in \fref{combined}(e).
The presence of various types of sharp peaks in the plot of HB density indicates the high probability of forming HBs between a particular combination of donor and acceptor pairs.
%
%
For instance, the promiment peak inside {\bf S}--{\bf W1} region in \fref{combined}(e) refers to the probability of donating HBs by {\bf S} layer to {\bf W1} layer.
Similarly, 
the presence of other peaks suggests that strong HB network is established between {\bf W1}--{\bf S} and {\bf W1}--{\bf W1} donor--acceptor combinations, whereas the probability of forming HBs between {\bf S}--{\bf W2} and {\bf W1}--{\bf W2} combinations is rather less. 

%
This HB network formation is closely related to the three dimensional structure of the interface and the orientation of the water molecules.
In~\fref{combined}(c),
we analyzed orientation angle ($\theta$) distribution of O--H bonds to get further clarification on the types of HB interaction at the interface. 
The surface O--H bonds, resulting from the protonation of surface O atoms, 
are oriented to upward direction
%
with orientation angle maxima at $\sim$69$^{\circ}$ and $\sim$11$^{\circ}$ as shown in \fref{combined}(b, c and d).
The peaks at $\sim$69$^{\circ}$ and $\sim$11$^{\circ}$ corresponds to a orientations where they can donate HB to water molecules of {\bf W1} and {\bf W2} layer to form {\bf S}--{\bf W1} and {\bf S}--{\bf W2} types of HB paterns, respectively. 
{\bf M}-type waters of {\bf W1} layer have nearly flat orientation with
%
orientation angle peaks at $\sim$85$^{\circ}$ and $\sim$99$^{\circ}$ (see \fref{combined}(b)).
They donate one HB ({\bf W1}--{\bf S}) to the surface O atoms and the corresponding
O--H bond is tilted downwards to the interface with orientation angle $\sim$99$^{\circ}$.
Meanwhile,
they donate another HB ({\bf W1}--{\bf W1})
to {\bf D}-type water present nearby in the same layer,
%
which corresponds to an orientation angle of $\sim$85$^{\circ}$.
The O--H bonds of the {\bf D}-type water molecules are
pointed upwards from the surface with orientation angle of $\sim$37$^{\circ}$,
%
donating HBs ({\bf W1}--{\bf W2} type) to {\bf W2} layer (see \fref{combined}(d)).
%
%
%
In contrast, the water molecules of {\bf W2} layer
are more free to rotate than the {\bf W1} layer and 
can  donate H-bond to {\bf S}, {\bf W1}, {\bf W2} and outer layers.
%
%
Notably, the peaks of the HB distribution plots are sharp and isolated from each others, clearly
indicating the presence of a well defined network of HBs at the interface.
%
Similar type of HB network was also found to be present in InP/GaP (001) surfaces in contact with liquid water.~\cite{Wood:JACS:2013}

Overall, a three dimensional HB network is formed at the interface,
opening up various channels of proton diffusion.
In particular, the presence of strong HBs with different donor--acceptor combination in the {\bf S} and {\bf W1} layer leads to {\bf W1W1} and {\bf SW1} paths through which proton can hop, utilizing the low barrier HB (see \fref{proton_transfer}).
The high rate of proton transfer in these channels is primarily due to the active intermediates being well connected by the HB network and easily converting to another form by accepting or donating a HB.




\section{Conclusion}
In summary, we have investigated the structure and dynamics of the (100) $\beta$--TaON--water interface under ambient conditions employing AIMD simulations.
%
Our observations reveal a high reactivity of the surface toward water, leading to the formation of various active intermediates, like HO$^-$, O$_{\rm S}$H and H$_3$O$^+$, resulting in a partially hydroxylated surface.
%
%
Nearly 44\% of the surface O atoms become protonated  as a result of water dissociation.
This protonation leads to the formation of a layered water structure perpendicular to the surface, with the highest water density at the interface.
%
%
%

At equilibrium, a well-defined HB network is established at the interface, enabling the spontaneous breaking and formation of surface chemical bonds such as O-H, Ta-OH, and Ta-Water.
This phenomenon facilitates rapid proton transfer processes on the surface.
Our study reveals multiple active routes of proton transfer at the interface, primarily driven by the presence of low-barrier HBs, 
which significantly enhances the rate of proton transfer at water--TaON interface.
While short-range proton transfer predominates at the surface, occasional diffusion of hydroxyl ions through multiple layers occurs via long-range proton transfer events.
Importantly, we find that the free energy barrier for proton transfer is very low, enabling rapid proton movement.
%
%

%
The presence of low-barrier HBs and multiple proton transfer
pathways could play a crucial role in proton diffusion during photo-catalytic water splitting reactions on TaON-based catalyst surfaces.
It has been reported previously that the high rate of proton transfer events could potentially enhance the photo-catalytic activity of InP–water interface.~\cite{Wood:JACS:2013}
Similarly, a higher proton transfer rate at TaON-water interface could reduce the
barrier for H$_2$ evolution reactions, favoring overall water splitting.
We believe that the overall water splitting activity on TaON-based catalyst surfaces is attributed to readily accessible proton at the active sites of the surface through fast proton diffusion mechanism, which could minimize electron and hole recombination. 
This has been realize in recent experiments using Ru/Cr$_2$O$_3$/IrO$_2$ co-catalysts on Zirconium-doped TaON, which shows a stoichiometric water splitting into hydrogen and oxygen is possible with an improvement in efficiency.~\cite{Xiao_AngewChem:2022}
We believe that the insights into the structural and dynamical properties of the water-TaON interface presented in this study provide a foundation for understanding the mechanisms of molecular catalysis on TaON surfaces in aqueous environments.

\begin{acknowledgement}

The authors thank Prof. Nisanth N. Nair (IIT Kanpur, India) for his support and encouragement to pursue independent research during their Ph.D. studies.
Computational resources were provided by the HPC facility at Indian Institute of Technology Kanpur.

\end{acknowledgement}




\bibliography{acs-achemso}

\end{document}